\documentclass[conference]{IEEEtran}
\IEEEoverridecommandlockouts
\usepackage{cite}
\usepackage{amsmath,amssymb,amsfonts}
\usepackage{algorithmic}
\usepackage{graphicx}
\usepackage{textcomp}
\usepackage{xcolor}
\def\BibTeX{{\rm B\kern-.05em{\sc i\kern-.025em b}\kern-.08em
    T\kern-.1667em\lower.7ex\hbox{E}\kern-.125emX}}
    
\usepackage{url}
\usepackage[linesnumbered,ruled,vlined]{algorithm2e}
\SetKw{KwTo}{in}

\begin{document}

\title{A Pub-Sub Architecture to Promote Blockchain Interoperability
\thanks{This project has been supported by \emph{The Linux Foundation} as part of the \emph{Hyperledger Summer Internships} program under the \emph{Towards Blockchain Interoperability with Hyperledger} project.}
}

\author{\IEEEauthorblockN{Sara Ghaemi\IEEEauthorrefmark{1}, Sara Rouhani\IEEEauthorrefmark{2}, Rafael Belchior\IEEEauthorrefmark{3}, Rui S. Cruz\IEEEauthorrefmark{3}, Hamzeh Khazaei\IEEEauthorrefmark{4}, Petr Musilek\IEEEauthorrefmark{1}}
\IEEEauthorblockA{\IEEEauthorrefmark{1}University of Alberta, Edmonton, Canada, \{sghaemi, petr.musilek\}@ualberta.ca}
\IEEEauthorblockA{\IEEEauthorrefmark{2}University of Saskatchewan, Saskatoon, Canada, sara.rouhani@usask.ca}
\IEEEauthorblockA{\IEEEauthorrefmark{3}Instituto Superior Técnico, Universidade de Lisboa, Lisboa, Portugal, \{rafael.belchior, rui.s.cruz\}@tecnico.ulisboa.pt}
\IEEEauthorblockA{\IEEEauthorrefmark{4}York University, Toronto, Canada, hkh@yorku.ca}
}

\maketitle

\begin{abstract}
The maturing of blockchain technology leads to heterogeneity, where multiple solutions specialize in a particular use case. While the development of different blockchain networks shows great potential for blockchains, the isolated networks have led to data and asset silos, limiting the applications of this technology. Blockchain interoperability solutions are essential to enable distributed ledgers to reach their full potential. Such solutions allow blockchains to support asset and data transfer, resulting in the development of innovative applications.

This paper proposes a novel blockchain interoperability solution for permissioned blockchains based on the publish/subscribe architecture.  We implemented a prototype of this platform to show the feasibility of our design. We evaluate our solution by implementing examples of the different publisher and subscriber networks, such as Hyperledger Besu, which is an Ethereum client, and two different versions of Hyperledger Fabric. We present a performance analysis of the whole network that indicates its limits and bottlenecks. Finally, we discuss the extensibility and scalability of the platform in different scenarios. Our evaluation shows that our system can handle a throughput in the order of the hundreds of transactions per second. 
\end{abstract}


\section{Introduction} \label{sec:intro}

The distributed ledger technology (DLT) enables a set of independent untrusted nodes to establish an agreement on the state of a shared ledger. Blockchain, a type of DLT, is mostly known for its use cases in cryptocurrencies such as Bitcoin~\cite{nakamoto2019bitcoin}, Ethereum~\cite{wood2014ethereum}, and XRP\cite{chase2018analysis}, among others. However, the technology can be used for more diverse applications and industries. Some examples are biomedical and health care~\cite{kuo2017blockchain,rouhani2018medichain}, Internet of Things (IoT)~\cite{fernandez2018review,fan2019towards}, public administration \cite{belchior2019,belchior2019_audits}, and cloud computing~\cite{ghaemi2020chainfaas,park2017blockchain}. Since each industry has its own unique sets of requirements, many isolated permissioned and permissionless blockchains have been introduced, posing limits regarding its interoperability.

Currently, developers commit to a single blockchain solution, and they cannot use the capabilities of more than one blockchain (vendor lock-in). These isolated, incompatible networks have resulted in silos of data and assets, which cannot be used from other networks. Blockchain interoperability solutions are needed to enable asset and information transfer from one blockchain to another. However, interoperability for blockchains encounters challenges that make it different from interoperability for other software networks. 

First, each interoperability solution should take into account the differences in the architecture of blockchain networks. Although all blockchains have an immutable ledger that stores the history of assets, they usually reach a consensus on the order of transactions using different algorithms. As a result, their underlying network and their validation mechanisms can be different from other blockchains. Each blockchain network that participates in the interoperation is independent and in full control of their assets and information. Moreover, the interoperability solutions should not require significant changes in the underlying blockchain networks, and it should be usable with minimal effort for existing blockchains.

This paper aims to tackle this problem by proposing a blockchain interoperability solution based on the publish/subscribe architecture across permissioned blockchains. 
We have implemented a broker blockchain that acts as a middleman in the interoperability process between the source and the destination networks. It is worth noting that since the broker is itself a blockchain network, it is not a central authority and peers from the source and destination blockchains can also participate in the governance of this network. The \emph{broker} blockchain keeps a copy of the information that needs to be shared in the form of a \emph{topic}. A topic has a name, message, publisher, and a set of subscribers. The \emph{publisher} is the source blockchain network that wants to share the information. It is responsible for creating the topic on the broker and publishing it to the corresponding topic whenever the information needs an update. The \emph{subscribers} are the destination networks that need some information from the source network. As soon as the subscriber network subscribes to a topic, the broker network notifies it whenever a change happens. This solution enables interoperability between blockchains with minimal effort. 

We used a performance benchmark tool to analyze the performance of the implemented prototype of the platform. The throughput and average latency for different functionalities of the broker network were investigated. The results indicate that our network can handle hundreds of transactions per second. Moreover, the evaluations identified the \textit{PublishToTopic} functionality to be the broker network's bottleneck.

The rest of this paper is organized as follows. Section \ref{sec:relatedWork} gives a summary of the related work on blockchain interoperability and blockchain-based publish/subscribe protocols. Section \ref{sec:sysDesign} introduces the system design details for the proposed interoperability solution. Section \ref{sec:implementation} demonstrates the implementation and deployment details of the platform, while section \ref{sec:evaluation} presents its performance evaluation. Section \ref{sec:discussion} outlines some discussions on the design and evaluation of the platform and section \ref{sec:conclusion} concludes the paper.

\section{Related Work} \label{sec:relatedWork}
In this section, we discuss the related work in the field of blockchain interoperability, as well as blockchain-based publish/subscribe protocols.

\subsection{Blockchain Interoperability}
The emergence of the blockchain interoperability research area has brought attention from both the industry and academia. Blockchain interoperability projects have been tackled as early as in 2016 \cite{belchior2020survey}.

A recent survey classifies blockchain interoperability studies in three categories: Cryptocurrency-directed interoperability approaches, Blockchain Engines, and Blockchain Connectors~\cite{belchior2020survey}. Cryptocurrency-directed approaches are mostly industry solutions that provide interoperability across public blockchains. This category has a focus on asset interoperability and is divided into sidechains, hash lock time contracts, notary schemes, and hybrid solutions. Sidechains allow for offloading transactions to a secondary chain, enhance performance, and provide features that the main chain would not provide. Sidechains also enable the representation of a token from the mainchain at the secondary chain. Some sidechain solutions include the BTC Relay~\cite{btcrelay}, Zendoo~\cite{zendo}, and RSK~\cite{rsk}. Hash lock time contract solutions enable cross-chain atomic operations using smart contracts. Wanchain uses this scheme and provides loan services with cryptocurrencies~\cite{Lu2017}. Notary schemes are centralized or decentralized entities that mediate token exchange (e.g., cryptocurrency exchanges). Finally, hybrid solutions combine characteristics from previous approaches. The cryptocurrency-directed approaches only work for transferring different types of cryptocurrencies between blockchain network. As a result, these approaches cannot be used for permissioned blockchains with arbitrary assets and smart contracts, which are the focus of this paper. 

The second category is blockchain engines, which enable creating customized blockchains that can interoperate by providing reusable data, network, consensus, and contract layers. Platforms like Polkadot~\cite{Wood2017} and Cosmos~\cite{Kwon2016} provide such infrastructure, with ``free interoperability'' among the instances they allow to create. These approaches are fundamentally different from what has been proposed in this paper. Instead of enabling blockchain interoperability for currently running blockchains, blockchain engines propose blockchain networks that are interoperable by design. As a result, these solutions cannot be used for currently running permissioned blockchains.

The Blockchain Connector category is composed of interoperability solutions that are not cryptocurrency-directed or blockchain engines. They include trusted relays, blockchain agnostic protocols,  blockchain of blockchains solutions, and blockchain migrators. Each of these categories responds to a particular set of use cases.
Trusted relays allow discovering the target blockchains, often appearing in a permissioned blockchain environment where trusted escrow parties route cross-blockchain transactions.  An interesting trusted relay approach is Hyperledger Cactus~\cite{cactus_whitepaper}, the most recent Hyperledger project aiming to connect a client to several blockchains, whereby transactions are endorsed by trusted validators. Cactus focuses on providing multiple use case scenarios via a trusted consortium. Another trusted relay is proposed by Abebe et al. \cite{abebe2019enabling}, which is an interoperability solution between Fabric networks using Hyperledger Fabric chaincode and a protocol-buffer based communication protocol. Blockchain agnostic protocols enable cross-blockchain communication between arbitrary distributed ledger technologies, including refactoring and making changes to each blockchain.  Blockchain of blockchains are approaches that allow users to build decentralized applications using multiple blockchains. Finally, blockchain migrators enable the state of one blockchain to be migrated to another blockchain. Simple blockchain migrations have already been performed, paving the direction for complex blockchain migrations~\cite{belchior2020survey}.


Our pub-sub solution is considered a trusted relay (although decentralized), as it contains a blockchain mediating communication across heterogeneous blockchains~\cite{belchior2020survey}. While trusted relays can provide a straightforward way of achieving interoperability, most of them are not trustless (e.g., contain a centralization point). Our solution is a decentralized trusted relay that implements a pub/sub system, anchored on the trust that underlying blockchains offer.

 
\subsection{Blockchain-based Publish/Subscribe Protocol}

The blockchain technology has been applied to the pub/sub paradigm in a few previous studies. However, those studies adopt blockchain to address the existing problems in other areas, such as IoT \cite{lv2019iot}, supply chain \cite{ramachandran2019trinity}, multi-tenant edge cloud \cite{huang2020bps}, and digital trading \cite{bu2019hyperpubsub}. 

Huang et al.~\cite{huang2020bps} exploit blockchain technology to enhance the security of pub/sub communications in multi-tenant edge clouds. Most topic-based and broker-enabled pub/sub streaming systems use centralized cloud servers to store sensitive metadata and access control lists (ACL), which can compromise the confidentiality, anonymity, and integrity of tenants' data. Alternatively, critical data such as ACL and identity information, as well as the hash of raw messages, and operation logs, can be stored on the blockchain to guarantee data security and integrity. Their smart contracts implement access control mechanisms to authorize requests from publishers and subscribers. 

Trinity~\cite{ramachandran2019trinity} proposes a blockchain-based distributed publish/subscribe broker to solve the existing flaws of centralized brokers in IoT and supply chain monitoring applications. Trinity has three main components: blockchain network, broker, and clients. The blockchain network is responsible for consensus in the system and persistent storage. The broker handles the communications between the blockchain network and clients. The clients are the publishers and subscribers of the topics. The blockchain network is pluggable, and the authors have used Tendermint, Hyperledger Fabric, Ethereum, and IOTA. The broker has used the Mosquitto MQTT broker, which provides a single point of failure.

Zhao et al.~\cite{zhao2018secure} have proposed Secure Pub-Sub (SPS), which provides fair payment based on a reputation for publishers and subscribers in cyber-physical systems. They use the Bitcoin's network to enable payments between the entities, and they propose a reputation mechanism that helps calculate the price of data.

Lv et al.~\cite{lv2019iot} presents a decentralized privacy-preserving pub/sub model for IoT systems to solve centralized brokers' problems such as single point of failure, data tampering due to corrupter brokers, and heavy encryption algorithms. The presented model applies public-key encryption with equality test \cite{yang2010probabilistic} and ElGamal \cite{elgamal1985public} to protect participants' (both publishers and subscribers) privacy. A system prototype is implemented and evaluated against the feasibility of the proposed model. 

Bu et al.~\cite{bu2019hyperpubsub} and Zupan et al.~\cite{zupan2017hyperpubsub} have proposed blockchain-based pub/sub brokers to address the drawbacks of traditional pub/sub systems such as privacy and accountability. However, they have not explained their implementation and evaluation in their studies. 

All these studies adopt blockchain to improve the centralized predicaments in traditional pub/sub systems in distinct application domains, whereas our paper focuses on establishing robust and practical interoperability between permissioned blockchains with different architectures and infrastructures. To the best of our knowledge, our paper is the first study that enhances blockchain interoperability utilizing the pub/sub communication model across heterogeneous blockchains (Hyperledger Fabric/Hyperledger Besu). 


\section{System Design} \label{sec:sysDesign}

In this section, we first discuss the design principles that a blockchain interoperability solution should follow. We then propose our interoperability solution and explain its components and message flow. 

\begin{figure*}[!htb]
    \centering
      \includegraphics[width=0.85\textwidth]{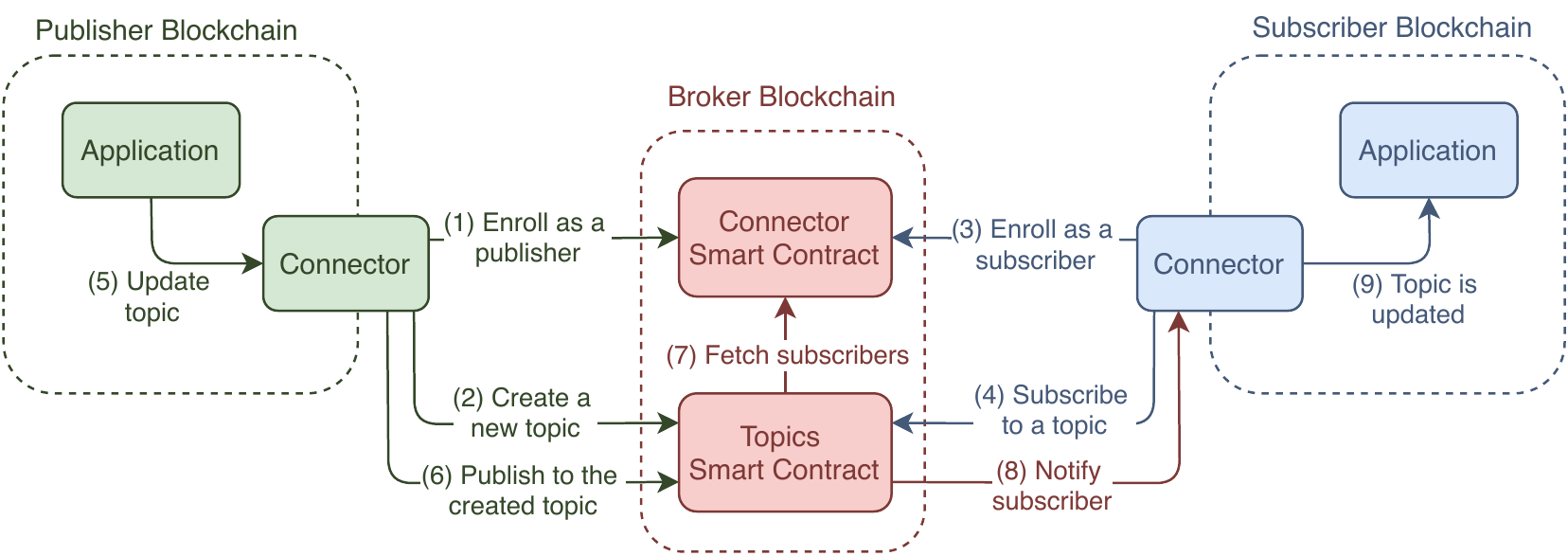}
      \caption{Architecture of the platform and the message flow.} 
      \label{Fig:Architecture}
\end{figure*} 

\subsection{Design Principles}

Blockchain interoperability aims to enable applications to use the assets and information available on blockchains other than their main blockchain network \cite{belchior2020survey}. This allows for a greater range of applications. A blockchain interoperability solution should take into account the following design principles:
\begin{itemize}
    \item The blockchain networks are independent, and they may have different architectures. 
    \item The blockchain networks are in full control of their assets and information.
    \item The transfer protocol should be technology agnostic. \item The interoperability solution should not require significant changes in the source and destination networks. 
    \item The blockchain networks should be able to incorporate the solution with minimal effort.
\end{itemize}

Following the mentioned principles, we present our solution, which allows interoperability based on a publish/subscribe architecture.

\subsection{Components}

The platform proposed in this paper aims to solve the interoperability problem of permissioned blockchains using the publish/subscribe pattern. When a blockchain wants to use the data from another blockchain, there needs to be a way to fetch and transfer this data between the networks securely. Moreover, if the data changes in the source network, the destination network should be notified of the change. Figure~\ref{Fig:Architecture} shows the architecture of the platform and the message flow. 

The \textit{publisher blockchain} is the blockchain network that sends the data, also referred to as the source network. For a publisher to participate in this platform, it needs to run the appropriate connector smart contract on its blockchain and enroll as a publisher in the broker. The publisher can then create as many topics as they want and use the connector smart contract to publish the changes to the topic. 

The \textit{subscriber blockchain} is the blockchain network that received the data, also referred to as the destination network. Similar to the publisher, the subscriber also needs to run the appropriate connector smart contract and enroll as a subscriber. It can then subscribe to any topic available on the broker blockchain. Every time the topic changes, the broker notifies the subscriber by invoking the connector smart contract. There can exist many subscribers for a topic.

The \textit{broker blockchain} is the core component of the platform. It is a blockchain network that stores all the information about the topics and the blockchains that participate in the interoperability process. Since the broker is a blockchain, operations regarding interoperation are immutable and traceable, providing a robust basis for accountability. The broker has two different smart contracts, the \textit{connector smart contract} and the \textit{topics smart contract}. The connector smart contract stores the information about the participating blockchain networks and how the broker can interact with them. The topics smart contract is responsible for storing the information about the topics such as their publisher, subscribers, and the latest message.

\subsection{Message Flow}

The complete interoperation process in the platform is shown in Figure~\ref{Fig:Architecture}. For simplicity, only one publisher and one subscriber blockchain are shown in this figure. However, for each topic, there can be an unlimited number of subscribers and, in general, there is no limit on the number of publisher and subscriber blockchains. A detailed explanation of each step in Figure~\ref{Fig:Architecture} follows:

\begin{enumerate}
    \item For any blockchain network to interact with the broker blockchain, it must enroll in the connector smart contract. During this process, the information that the broker needs to be able to interact with the blockchain is registered in the connector smart contract. As a result, the publisher is required to enroll in the connector smart contract as a publisher. This step only needs to be performed once, when the publisher wants to create its first topic. It can then create topics or publish to existing ones without the need to be enrolled again.
    \item A publisher that is registered in the connector smart contract can create a new topic. In this step, the publisher needs to specify a name for the topic and the initial topic message. This step only needs to be executed once for each topic.
    \item Similar to the publisher blockchain, the subscriber blockchain should also enroll in the connector smart contract. This step only needs to be done once, when the subscriber wants to subscribe to a topic for the first time.
    \item To receive a notification when a topic has been changed, the subscriber should subscribe to the topic by sending a request to the topics smart contract. This results in the subscriber being added to the list of all subscribers for the topic. This step only needs to be performed once for each topic. 
    \item Whenever needed, the application in the publisher blockchain can update the topic by sending a request to the connector smart contract. 
    \item The connector smart contract sends a publish request to the topics smart contract for the existing topic. 
    \item As soon as a publish request is received by the topics smart contract, the smart contract fetches the information about all the subscribers of the topic from the connector smart contract. It includes information on how the broker can interact with each of the subscribers. 
    \item The topics smart contract then uses the data fetched from the connector smart contract to notify all the subscribers of the change in the topic. This happens by sending an update request to the connector smart contract in each of the subscriber networks. 
    \item In each subscriber network, the connector smart contract receives the update for the topic and notifies the application.
\end{enumerate}

\section{Implementation and Deployment} \label{sec:implementation}

A prototype of the proposed platform has been implemented as a proof-of-concept to demonstrate the feasibility of the design. This project is a Hyperledger Labs open-source project\footnote{\url{https://github.com/hyperledger-labs/pubsub-interop}}, under the Apache License, Version 2.0. 
To show the interoperability capabilities of the platform, we implemented two example subscriber networks, as well as an example publisher network. The broker and the example publisher are implemented using two different Hyperledger Fabric V2~\cite{fabric2Docs} networks. The two example subscribers are implemented using Hyperledger Fabric V1.4~\cite{fabric1Docs,androulaki2018hyperledger}, and Hyperledger Besu~\cite{besuDocs}. In this section, the implementation and deployment details of the broker and the example networks are discussed. 

\subsection{Broker Blockchain}

The broker blockchain acts as a messaging broker between other blockchains to enable interoperability. When choosing the blockchain solution to implement the broker network, we had to ensure that the solution fits well with the needs of this platform. First, since we aim to address interoperability in permissioned blockchains, the broker also needs to be permissioned. Moreover, many permissioned blockchains are enterprise-level, and they may have privacy and governance concerns, which our broker aims to addresses using blockchain that enables immutability and traceability. We need a blockchain solution for the broker network that considers these needs. Finally, the smart contracts that need to be implemented on the broker blockchain are complicated, and the blockchain needs to support this kind of smart contract. 

Hyperledger Fabric is an open-source permissioned blockchain that has been designed for enterprise use cases. Unlike the open permissionless blockchains that have scalability issues, Fabric enables high transaction throughput and low transaction confirmation latency. The architecture of Hyperledger Fabric is highly modular and configurable, which enables customization for each specific use case. Many blockchains only support smart contracts written in non-standard and domain-specific programming languages, making it impossible to implement smart contracts that require a Turing-complete language. On the other hand, Hyperledger Fabric supports smart contracts in general-purpose programming languages such as Go, Node.js, and Java~\cite{fabric2Docs}.

In the broker network, we leverage the capabilities of Hyperledger Fabric V2.2 to implement a messaging broker. We implement two chaincodes called the topics and the connector to support the features needed for the broker. The topics chaincode is responsible for keeping all the topics and their corresponding details. In Hyperledger Fabric, everything is stored as a key-value pair. In the topics smart contract, the key is a unique topic ID. The value is an object that includes the following properties: name, publisher, subscribers, and message. The name of a topic is a string value set by the publisher when creating the topic. Each topic has one publisher, the blockchain network that has registered the topic on the broker, which is the only blockchain that can make changes to the topic. The subscribers' property stores a list of all the blockchains that have subscribed to the topic. It is worth mentioning that the publisher and the subscribers' properties only accept objects stored on the connector blockchain. 

The connector chaincode is responsible for storing the connection details of other blockchain networks. Similar to the topics chaincode, the key in the key-value pair used in this chaincode is a unique ID for each blockchain. The value is an object that has the following properties: name, type, server IP, port, extra information. The name is a string value that can be selected when enrolling in the network. Type shows what kind of blockchain technology this network is using. Currently, support for Fabric and Besu has been implemented, and other blockchains will be added in future versions. The server IP and port are used by the broker blockchain to access the publisher or subscriber using an HTTP request. The extra information property stores network-specific details that may be needed when interacting with the blockchains. For instance, for a Hyperledger Besu network, this includes the private key, address, and the contract application binary interface (ABI) that the broker should use to send a request to the Besu network. This kind of implementation allows our solution to be extendable to other blockchains, independent of their underlying network. 

\begin{figure}[!htbp]
    \centering
      \includegraphics[width=0.85\columnwidth]{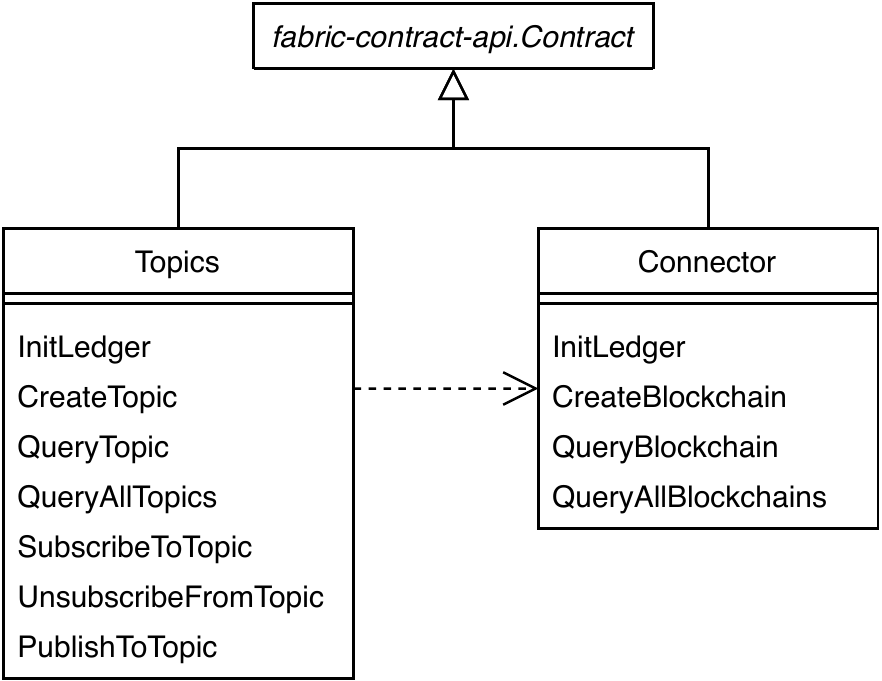}
      \caption{UML class diagram of the implemented chaincodes.} 
      \label{Fig:ClassDiagram}
\end{figure}

To better understand how the topics and connector chaincodes work, we need to discuss their implemented functionalities. Figure~\ref{Fig:ClassDiagram} shows the UML class diagram of the implemented chaincodes. The Hyperledger Fabric contract API provides an interface for developing smart contracts and applications. Each developed chaincode should extend the contract class from this API and then implement the required logic. In each smart contract, the \textit{InitLedger} function is used to create a set of initial assets on the ledger when the chaincode is deployed. In the topics chaincode, the \textit{CreateTopic} function is used to create a new asset of type topic. The \textit{QueryTopic} and the \textit{QueryAllTopics} functions can be used to query one specific topic and all the existing topics, respectively. The connector chaincode implements the same initialize, create, and query functionalities but for assets of type blockchain. 

Other than the mentioned functions, the topics blockchain also implements \textit{SubscribeToTopic}, \textit{UnsubscribeFromTopic}, and \textit{PublishToTopic} functionalities. When a destination blockchain wants to get notified of a topic's change, it subscribes to that topic by invoking the \textit{SubscribeToTopic} function. 
The subscriber can also unsubscribe from the topic at any time by invoking the \textit{UnsubscribeFromTopic} function. Finally, the \textit{PublishToTopic} function is used by the source blockchain network when they want to update the topic's message. An invoke request to this function triggers update requests to all the subscribers of the topic. Algorithm~\ref{algorithm:publish} shows the detailed implementation of the \textit{PublishToTopic} method. First, the broker retrieves the latest version of the topic from the ledger. In the case that no topic was found, it immediately throws an error. Next, the topic's message is updated with the new message value and the topic's state is put to the ledger. The next step is for the broker to notify all the subscribers. For each of the subscribers of the topic, the blockchain object is queried from the connector contract. This inter-chaincode communication is also shown in Figure~\ref{Fig:ClassDiagram}. Then, given the type of subscriber blockchain, the steps to invoke the remote network are followed. 

\begin{algorithm}[!htbp]
\small
\SetAlgoLined
\KwIn{topicID, newMessage}
    \KwResult{Subscribers are notified of the new message}
topicState $\gets$ getState(topicID)\\
\If{!topicState}{
  throw error\\
 }
topicState.message $\gets$ newMessage\\
putState(topicID, topicState)\\
\For{subID \KwTo topicState.subscribers }{
    subState $\gets$ query $subID$ from connector contract\\
    \uIf{subState.type = Fabric}{
        follow steps to invoke a remote Fabric network
    }
    \ElseIf{subState.type = Besu}{
        follow steps to invoke a remote Besu network
    }
    }
 \caption{PublishToTopic Method}
 \label{algorithm:publish}
\end{algorithm}

\subsection{Subscriber Blockchains}

The subscriber, or destination blockchain, is the blockchain that requires information from another blockchain to run a task. For the subscriber to be able to participate in the platform, it needs to have the appropriate connector smart contract deployed on it. We have implemented subscriber connector contracts for Hyperledger Fabric V1.4 and Hyperledger Besu. However, the connector is a simple smart contract that can also be developed by the owners of the subscriber blockchain. This smart contract needs to keep track of the topics that the subscriber has subscribed to and store their latest version for other smart contracts to access at any time. Two example subscriber networks have been implemented to demonstrate the interoperability capabilities of the platform. 

The first example subscriber is implemented using Hyperledger Fabric V1.4. The second example subscriber is implemented using Hyperledger Besu, an open-source Ethereum client that supports private and permissioned blockchains. Besu can create networks that work based on a proof of work (PoW) or a proof of authority (PoA) consensus algorithm. In this work, we implemented a PoW network using Besu, which can be thought of as a private Ethereum network. We then implemented a connector smart contract in Solidity to keep a record of the subscribed topics.



\subsection{Publisher Blockchains}

The publisher, or the source blockchain, is the blockchain network that needs to send information to other blockchains. Similar to what we have in the subscriber blockchain, a connector smart contract is also required for the publishers. However, the connector is slightly different in the publisher. The publisher connector should not only keep track of the topics, but it should also connect to the broker blockchain to publish the topics. We implemented an example publisher network using Hyperledger Fabric V2.2.

\section{Experimental Evaluation} \label{sec:evaluation}




In this section, we focus on evaluating the performance of the implemented prototype of the broker blockchain. The goal is to see how the throughput and latency of the system changes in different scenarios. We have conducted two series of experiments to achieve this goal. The first set of experiments aims to show the performance metrics of different functionalities in the broker blockchain. The second set of experiments focuses on the publish function, which is the most important and time-consuming component of the broker blockchain. 

We have used Hyperledger Caliper~\cite{hyperledgerCaliper} to run the experiments. Hyperledger Caliper is an open-source blockchain performance benchmark tool that allows performance measurement for different blockchains, such as Hyperledger Fabric, Ethereum, Hyperledger Besu. In Hyperledger Caliper, the workloads or benchmarks are responsible for generating the content of each transaction that is sent to the blockchain network. Given the network and benchmark configurations, Caliper uses a set of independent workers to send scheduled requests to the blockchain network and monitor the response. When the tests are finished, Caliper generates a performance report consisting of the average throughput and minimum, maximum, and average latency throughout the test. The throughput shows the number of transactions that were processed in the system in a given time. The latency shows the amount of time it takes for a transaction to be finished and added to the ledger. 

Table \ref{tab:experimental_setup} summarizes the specifications of each component in the experimental evaluation. We have set up Hyperledger Caliper on a separate machine to ensure that its process does not affect the performance of the broker network. We use five workers, a fixed rate controller, and a test duration of 60 seconds for each benchmark round.
The broker network is implemented using Hyperledger Fabric V2.2 with two peer organizations and an orderer organization, each with an independent certificate authority. Each of the peer organizations hosts one peer node, and the orderer uses Raft implementation. Two chaincodes have been implemented that run on one channel. The Fabric subscriber, implemented using Fabric V1.4, has two organizations, each hosting two peers. The whole subscriber network uses one Solo orderer and one Fabric certificate authority. The Besu subscriber implements a private Ethereum network with PoW consensus algorithm. The publisher has been implemented using Hyperledger Fabric V2.2 with the same configurations as the broker network.

\begingroup
\renewcommand{\arraystretch}{1.5}
\begin{table}[h]
    \centering
  \caption{Experimental evaluation setup}
  \label{tab:experimental_setup}
  \begin{tabular}{lcccc}
    \hline
     Component & Type & CPU & RAM & Disk\\
    \hline
    Caliper Benchmark &   N/A  & 8 vCPU   & 30 GB &   288 GB\\
    Broker   & Fabric V2.2 & 8 vCPU &  30 GB  &   288 GB\\
    Fabric Subscriber   & Fabric V1.4 & 2 vCPU &  7.5 GB  &   36 GB\\
    Besu Subscriber   & Besu & 2 vCPU &  7.5 GB  &   36 GB\\
    Publisher   & Fabric V2.2 & 2 vCPU &  7.5 GB  &   36 GB\\
  \hline
\end{tabular}
\end{table}
\endgroup

The first set of experiments focuses on the performance evaluation of broker blockchain. In these experiments, we conduct a series of tests using Hyperledger Caliper for each functionality that broker blockchain offers. Figure~\ref{Fig:ClassDiagram} summarizes all these functionalities. Each type of transaction goes through a specific set of steps in Hyperledger Fabric, which highly influences the response time for that transaction. For instance, an invoke transaction goes through endorse, order and commit steps. On the other hand, a query transaction is not transferred to the orderer, and the response is immediately sent back by the peer. The create actions in the connector and topics smart contract are invoke actions that have very similar implementations. The same goes for the query actions in the two smart contracts. As a result, it would be repetitive to run performance evaluation experiments for both smart contracts. Therefore, we run the experiments on the topics smart contract.

\begin{figure}[!tbp]
    \centering
      \includegraphics[width=0.95\columnwidth]{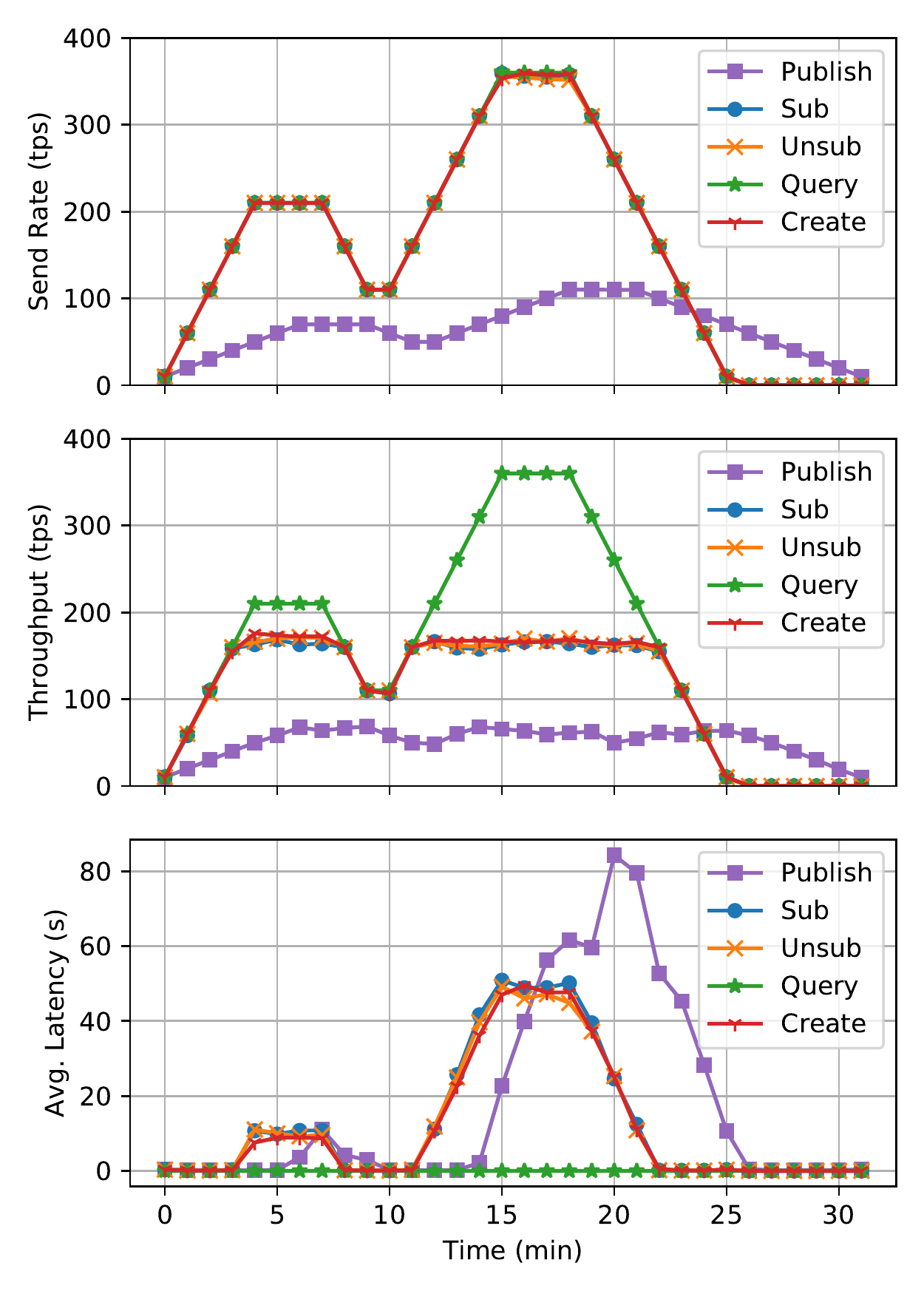}
      \caption{The trend of system throughput and average latency for various functionalities throughout time with the change of request send rate. The words publish, sub, unsub, query, and create in the plots stand for PublishToTopic, SubscribeToTopic, UnsubscribeFromTopic, QueryTopic, and CreateTopic functions, respectively.} 
      \label{Fig:EvalAll}
\end{figure}

\begin{figure}[!htbp]
    \centering
      \includegraphics[width=0.95\columnwidth]{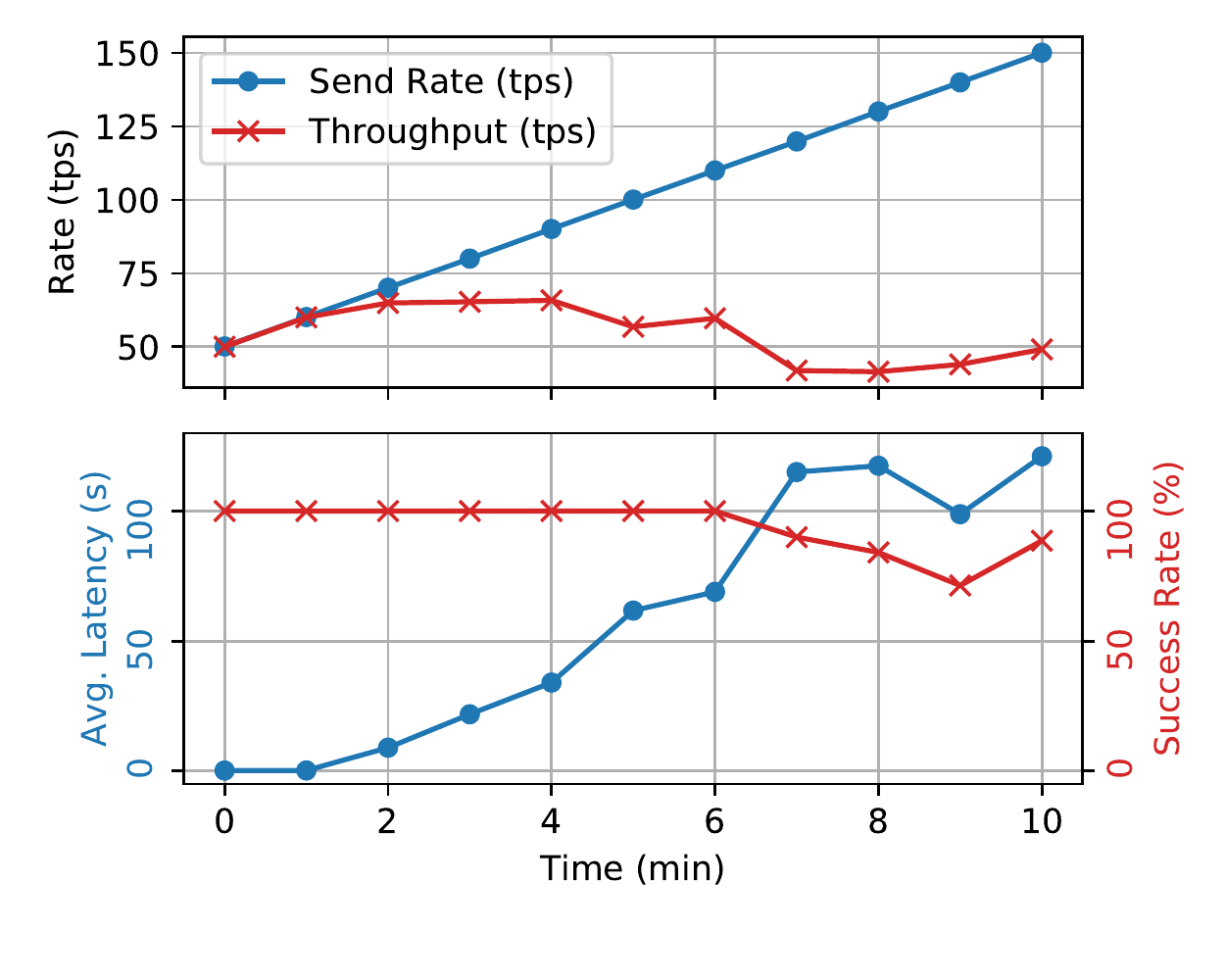}
      \caption{The trend of system throughput, average latency, and request success rate throughout time with the change of send rate.} 
      \label{Fig:EvalPub}
\end{figure}

The topics smart contract has five important functionalities: create a topic, query a topic, publish to a topic, subscribe to a topic, and unsubscribe from a topic. For each of these actions, we run a set of experiments by changing the transaction send rate in the Hyperledger Caliper benchmark. The goal is to see how the system's throughput and average latency changes when the send rate is changed. Figure~\ref{Fig:EvalAll} shows the details of these experiments. It can be seen that the send rate follows the same pattern for all the actions except for \textit{PublishToTopic}. The reason for this difference is that the \textit{PublishToTopic} action takes more time and needs more resources to run compared to other actions. Consequently, broker blockchain's hardware limits are reached when the network receives more than roughly 100 publish transactions in each second. We discuss the behaviour of the network with different \textit{PublishToTopic} requests in the second set of experiments shown in Figure~\ref{Fig:EvalPub}. As a result of this limitation, we lowered the send rate for the \textit{PublishToTopic} action in our experiments.

It can be seen in Figure~\ref{Fig:EvalAll} that the \textit{SubscribeToTopic}, \textit{UnsubscribeFromTopic}, and \textit{CreateTopic} have similar behaviours under the same send rate. These three actions are of type invoke. 
Since an invoke transaction proposes a change in the blockchain, it needs to go through the consensus algorithm, which can be time-consuming. Since the three actions are of the same type, and none need heavy computations in execution, the system throughput and latency for all of them are similar. As shown in the experimentation results, when the send rate is lower than a threshold (around 160 TPS in this case), the throughput is the same as the send rate, and the average latency is only a few hundred milliseconds (around 100 to 300 milliseconds). This shows that with send rates below the threshold, all transactions are processed immediately. When the number of create, subscribe, or unsubscribe transactions sent in each second is more than the threshold, the broker network's processing limit is reached. The throughput is limited to the broker's maximum capacity (around 160 TPS), and the transactions are queued before being processed, which results in an increase in the latency. Figure~\ref{Fig:EvalAll} shows that when the send rate for the create, subscribe, or unsubscribe transactions is around 210 TPS, the average latency increases to about 11 seconds. The latency keeps increasing with higher send rates and reaches approximately 50 seconds with a send rate of 360 TPS.

The \textit{QueryTopic} action is different from the previous ones. Since a query transaction does not go through the consensus protocol, its process is much faster. The send rate pattern used for the query is similar to that of create, subscribe, and unsubscribe. However, the throughput and average latency act very differently. The throughput follows the same pattern as the send rate, and the average latency is around 10 milliseconds throughout the whole experiment. These results show that this experiment does not reach the process limit for \textit{QueryTopic}. 

Finally, the \textit{PublishToTopic} is similar to create, subscribe, and unsubscribe because they are all invoke transactions. However, the publish action requires stronger computations. 
As mentioned earlier, since the publish action needs more time and computational resources, we use a different send rate pattern. If we were to use the same send rate, the broker blockchain's hardware limits would be reached, resulting in the experiments being halted. We discuss this in more detail in the second set of experiments shown in Figure~\ref{Fig:EvalPub}. To ensure that the performance of the remote source and destination networks do not influence the performance evaluation of the broker network, we only send dummy requests to the subscriber networks during the experiments. It can be observed from Figure~\ref{Fig:EvalAll} that the publish action reaches the processing limit of the broker network much faster than the other invoke transactions. With send rates of about 70 TPS and more, the throughput is limited to 65 TPS. The average latency for the publish action has more fluctuations compared to other invoke actions. The main reason for this fluctuation is that in the publish method, depending on the number of subscribers that the topic has, the processing time can vary. In this experiment, the average latency gets as high as 80 seconds, with the send rate of 110 TPS.

Given the limits of the \textit{PublishToTopic} action, we decided to run some additional experiments on this type of transaction. This experiment aims to find the broker network's limits and discover what happens when the limit is reached. In the previous experiment, we discovered that the processing limit for the publish transactions is reached at the sent rate of around 70 TPS. We also observed that the latency increases and the throughput is limited for send rates above 70 TPS and below 110 TPS. However, we would like to know what happens if the send rate is more than 110 TPS. In this experiment, we linearly increase the send rate from 50 to 150 TPS and observe the throughput, latency, and transaction success rate. Figure~\ref{Fig:EvalPub} shows the results of this experiment. Similar to the previous experiment, we see that the throughput is limited, and the latency is increased when the send rate reaches 70 TPS. Nevertheless, the interesting change happens at the 120 TPS send rate. At this point, a significant drop in the throughput and a significant rise of latency are observed. Moreover, the transaction success rate is not 100\% anymore. From this point on, a portion of the transactions fail since the broker network has reached its hardware limits.

\section{Discussion and Future Work}
\label{sec:discussion}

To enable blockchain interoperability, we have proposed the use of a broker blockchain as a middleman. The broker blockchain acts as a decentralized trusted relay between the source and destination network. Using a relay enables the interoperating networks to transfer data with minimal effort. Verifying the data and handling the communications between different blockchain networks can be delegated to the relay. As a result, there is no need for source and destination networks to make fundamental changes to their underlying structure. The relay network is also a blockchain network; while exploiting all desirable features offered by blockchain, it runs smart contracts implementing the interoperability functionality. Therefore, the broker blockchain allows the interoperation to be seamless, transparent, and secure.

The platform proposed in this paper stores the destination and source blockchains as assets on the distributed ledger. As a result, a large number of blockchains can be supported as there are no limits on the number of assets. A study on the performance of Hyperledger Fabric V1.1 shows that the network can scale to 100+ peers~\cite{performance2020Fan,Hyperledger2018Androulaki}. As the broker network has been implemented using Fabric V2.2, we expect this number to be even higher in our network. Therefore, at least 100 peers can participate in the governance of the broker blockchain. 


In the current prototype of our platform, every participant can subscribe to all existing topics, and there is no access control mechanism implemented. The private data feature presented by Hyperledger Fabric can be utilized to establish private channels in the broker blockchain and keep data topics separate from other organizations. 
Furthermore, an access control mechanism can be added to our pub/sub system to control the flow of data at a more granular level, for instance, the decentralized access control proposed by Rouhani et al.~\cite{rouhani2019}.

It is also possible to conduct authorization processes with minimal information disclosure if one uses the novel self-sovereign based access control model~\cite{ssibac}. This way, we could model each blockchain as an agent to prove that they own specific credentials needed for the access control process. 
Moreover, the publisher network may choose only to make a topic available to a subset of subscribers. Access control can be used to manage the blockchains that can access each topic.

\section{Conclusion} \label{sec:conclusion}

With blockchain technology gaining popularity in academia and industry, many blockchain networks are being introduced worldwide. These networks are highly isolated and incompatible with each other, resulting in silos of data and assets. Blockchain interoperability solutions can revolutionize this technology by enabling data and asset transfers between homogeneous and heterogeneous blockchains. In this paper, we proposed a blockchain interoperability solution based on the publish/subscribe architecture. Our solution consists of a broker blockchain that keeps a record of the data being transferred between blockchain networks. The blockchains that aim to participate in the interoperability can connect to the broker network as publishers or subscribers, depending on their role. A prototype of the broker blockchain has been implemented using Hyperledger Fabric. Moreover, an example publisher and two example subscribers have been implemented using Hyperledger Besu and two versions of Hyperledger Fabric to show that the design works for heterogeneous blockchains. The network's performance has been analyzed using a benchmark tool to identify the platform's limits and bottlenecks. The implementation and evaluations indicate the feasibility of the idea with satisfactory performance, and the bottleneck is identified to be the process of publishing a new message to a topic. Finally, a discussion on the extensibility, scalability, and possible improvements of the system is presented.

\bibliographystyle{ieeetr}
\bibliography{bibliography}

\end{document}